\documentstyle[aps,prl,epsf,twocolumn]{revtex}
\begin{document}
\def\bea{\begin{eqnarray}}
\def\eea{\end{eqnarray}}
\def\a{\alpha}
\def\d{\delta}
\def\p{\partial} 
\def\nn{\nonumber}
\def\r{\rho}
\def\rv{\bar{r}}
\def\la{\langle}
\def\ra{\rangle}
\def\e{\epsilon}
\def\o{\omega}
\def\n{\eta}
\def\g{\gamma}
\def\break#1{\pagebreak \vspace*{#1}}
\def\f{\frac}
\twocolumn[\hsize\textwidth\columnwidth\hsize\csname
@twocolumnfalse\endcsname
\draft
\title{Measuring the general relativistic curvature of wave-fronts}
\author{Joseph Samuel}
\address{Raman Research Institute,
Bangalore 560080, India\\}
\date{\today}
\maketitle
\widetext
\begin{abstract}
Einstein's general theory of relativity predicts that an initially 
plane wave-front will curve because of gravity.
This effect can now be measured using Very Long Baseline
Interferometry (VLBI). A wave-front from a distant point source 
will curve as it passes the gravitational field of the Sun.
We describe an idealised experiment to  directly 
measure this curvature, using four VLBI stations 
on earth, separated by intercontinental
distances. Expressed as a time delay, the size of the effect  
is a few hundred picoseconds and may be measureable 
with present technology.
\end{abstract}

]
\narrowtext
It is now possible to do interferometry with intercontinental baselines
(VLBI) and clocks of picosecond 
accuracy\cite{fomalont,sovers}. This technological
advance of the last few years suggests a new test of general relativity.
The test consists of directly measuring the curvature of 
a wave-front coming from a distant radio source. In the absence of 
general relativistic (GR) effects, the wave-front from the distant source 
would appear plane.
General relativity predicts that the wave-front will curve because
of gravitational effects. By means of four VLBI stations
located on the earth, one can directly measure this curvature.
The effect, whose size we estimate below, may be measurable with
currently available VLBI techniques. We describe an idealised
version of the experiment in the hope of motivating VLBI astronomers
to design more realistic experiments to detect the curvature of wave
fronts.

That gravity curves wave-fronts is already established from
astronomical observations of gravitational lensing. Multiple images,
caustics and ``luminous arcs'' prove that 
initially plane wave-fronts curve under the influence of gravity. 
However, all these 
gravitational lenses are outside the solar system and we can only
guess at and imperfectly model their structure. 
Tests of general relativity {\it within the solar system}
are far more under our control, since we are on familiar ground.
We can model the lenses in detail,
work out the predictions of the theory and 
{\it quantitatively} confront theory with experiment.

Two general relativistic effects in the propagation of light have already 
been accurately measured--the Shapiro time delay \cite{shapiro} 
and the bending of light\cite{will,will1}. 
It is worth noting that analogous effects exist even in special relativity.
The bending of light manifests itself as a change in the
apparent direction of the source. 
In special relativity, uniform motion of the
observer can result in aberration-an apparent change in the position
of the source. Time delays too can be induced by uniform motion of
the observer as in the Doppler effect.
However curving of plane wave-fronts is a
purely general relativistic effect that has no special relativistic
analog. A plane wave in one inertial frame
appears plane in all inertial frames.
The experiment proposed in this paper consists of directly
measuring the curvature of a wave-front, a purely 
GR effect with
no special relativistic analog.  

This letter is organised as follows: we first describe the theory behind
the proposed experimental test. We then describe the experiment 
in an idealised form and show how to extract the general relativistic
effect from measured quantities. 

Consider a plane wave incident on
an isolated static spherically symmetric body (the sun) of mass $m$ and 
radius $R$. We note that that 
$\epsilon:=2Gm/(c^2R)=2.10^{-6}<<1$ and 
neglect terms of order $\epsilon^2$.
The effect of gravity can be calculated using the Schwarzschild metric.
Neglecting terms of second and higher order in $\epsilon$, the 
Schwarzschild
metric is given in standard co-ordinates by 
(we henceforth set $c=1$).
\begin{equation}
ds^2=(1-\rho/r)dt^2-(1+\rho/r)dr^2-r^2d\Omega^2,
\label{metric}
\end{equation}
where $\rho=2Gm$ is the Schwarzschild radius of the body.
We can write the metric 
$g_{\mu\nu}=\eta_{\mu\nu}+h_{\mu\nu}+O(\epsilon^2)$,
where $h_{\mu\nu}$ is the General Relativistic perturbation of the metric 
and our notation $O(\epsilon^2)$ means that we neglect quantities which
are
second order and higher in $\epsilon$. 
The propagation of electromagnetic waves \cite{scalar} can be
described in the geometrical optics approximation by the eikonal
equation \cite{landau}
$$g^{\mu\nu}\partial_\mu\psi\partial_\nu\psi=0$$
 and the solution $\psi(x)$ is the phase of the wave, a scalar function of 
the general co-ordinate $x^\mu,\mu=0,1,2,3$.
Let us choose Minkowskian coordinates $x^\mu=(t,{\vec {\bf r}})$ adapted
to the flat metric
$\eta_{\mu\nu}$. Let $\omega$ be the frequency at infinity of the 
incident plane wave and let $k^M_\mu$ be  
wave vector at infinity: 
$k^M_{\mu}=(\omega,0,0,-\omega)$, where we have chosen the
direction of the incident plane wave along the positive
$z$ axis. In Minkowski 
space the 
function
$\psi^M(x)=k^M_{\mu}x^\mu$ with $\eta^{\mu\nu}k^M_\mu k^M_\nu=0$ solves 
the   
eikonal equation.
To first Post-Minkowskian order, the solution is
$\psi=\psi^M+\phi$, where $\phi$ satisfies the differential equation
\begin{equation}
\eta^{\mu\nu}k^M_\mu\partial_\nu\phi=1/2h^{\mu \nu}k^M_\mu k^M_\nu.
\label{perturb}
\end{equation}
Then (\ref{perturb}) leads to 
\begin{equation}
\phi(x,y,z)=-\frac{\rho \omega}{2} \int_{z_{-}}^z 
du[\frac{1}{f^{1/2}}+\frac{u^2}{f^{3/2}}],
\label{solution1}
\end{equation}
where $f=u^2+x^2+y^2$ and  
$z_{-}\longrightarrow-\infty$.  Integration yields:
\begin{equation}
\phi=\rho\omega\ln[(r-z)+z/2r]\Big{|}^z_{z_{-}}
\label{solution2}
\end{equation}
where $r=\sqrt{x^2+y^2+z^2}$. 
For lines of sight close to the sun
($x,y<<z$) it is more illuminating to use the 
simple approximate form
\begin{equation}
\phi=\frac{(1+\gamma)\rho\omega}{2}[\ln(x^2+y^2)/(2z)]
\label{solutionfin}
\end{equation}
which we  arrive at using a binomial expansion and dropping a constant
independent of $(x,y,z)$.
In (\ref{solutionfin}),
we have introduced the PPN 
(parametrised post-Newtonian\cite{will1}) parameter $\gamma$, which
is equal to $1$ in general relativity.

The total phase of the wave is given by
\begin{equation}
\psi=k^M_\mu x^\mu+\phi,
\label{phase}
\end{equation}
where $\phi$ is the small general relativistic correction, the Shapiro
delay.
The arrival of a wave-front at $(x,y,z)$ is delayed by a time 
$\phi(x,y,z)/\omega$ due to gravitational effects. Thus the gravitational 
field
acts as a retarder, just as a glass slab does with light. 
There is a close mathematical relation between the Shapiro delay
\cite{shapiro}, the bending of light \cite{will} and the curvature of
wave-fronts which is the subject of this paper.
Let us expand the eikonal $\psi$ in a Taylor expansion about an event 
$x_0$:
\begin{eqnarray}
\psi(x^\mu)&=& \psi(x_0)+(x^\mu-x_0^\mu) \partial_\mu 
\psi|_{x_0}+\nonumber\\ 
&&1/2(x^\mu-x_0^\mu) 
(x^\nu-x_0^\nu) \partial_\mu \partial_\nu \psi|_{x_0}+...
\label{taylor}
\end{eqnarray}
The Shapiro delay measures
$\psi(x_0)$, the eikonal. The bending of light
is a measurement of its first derivative $k_\mu=\partial_\mu \psi|_{x_0}$ 
and the
curvature effect described here measures its second derivative.
While these effects are related mathematically, they are distinct 
physical effects and should therefore be separately measured and checked 
against theoretical predictions. In these three effects the
gravitational field acts respectively as a retarder, a prism and
a lens, which act by delaying, tilting and curving a wave-front.

In the experiment to be described below, all observation points are on the 
earth (see figure 1). As  a result the numerical value of $h_{\mu \nu}$ 
the perturbation
of the metric tensor is around $10^{-8}$. 
We will therefore neglect $h_{\mu\nu}$ at the earth's
location and use the Minkowskian metric $\eta_{\mu\nu}$ to raise and 
lower indices. Henceforth all dot products are formed using the flat
Minkowskian metric. 

A wave-front is a three dimensional surface in four dimensional 
space-time. For easy visualisation let us pick a slice $t=0$.
The wave-front is now a two dimensional surface 
described by
\begin{equation}
-{\bf k}.{\bf x}+\phi({\bf x})=constant.
\label{twod}
\end{equation}
In order to detect the curvature \cite{curve} of this wave-front
one needs to sample at least four points on it. For, given
three points (or less), one can always find a plane passing
through all of them. The deviation of the fourth point
from the plane passing through the first three gives a measure
of the curvature of the wave-front. More symmetrically, one can compute
the volume of the tetrahedron with these four points as vertices. A 
non-zero volume would imply that these four points do not lie in a plane.
This simple three dimensional 
argument provides the intuition we will use in describing the experiment.
The principle of the experiment is that given the arrival 
events at three VLBI stations, one can model the
incident wave as a plane wave and absorb any deviation in a redefinition
of the apparent direction of the source. However, with the fourth station
this freedom does not exist and one can measure genuine 
curvature effects \cite{astro}. 
Our method does not require an absolute determination of the direction of 
the source.

Consider four VLBI stations ${\cal I}_a, a=0,1,2,3$ with intercontinental 
separations viewing a radio source along a line of sight 
passing near the sun. The VLBI experiment
involves simultaneously observing a structureless point radio source
from four telescopes.
The signal voltage received at each antenna is beaten with 
an ultra-stable local oscillator, low pass filtered, digitized 
and recorded. The recorded data from two antennae are cross-correlated 
with a delay to find the `fringe'-- the delay at which the 
cross-correlation peaks. Thus one can accurately locate the events
at which a given wave-front arrives at the four antennae. 

Such VLBI techniques accurately measure the arrival times $t_a$  of a 
wave-front
at each of these stations located at ${\bf x}_a$. Here $x_a^\mu=(t_a,{\bf 
x}_a)$ are the four four-vectors used 
to locate 
the 
four events, the arrival of a wave-front at each of the four stations. 
We will now describe how the measured quantities 
$x_a^\mu, a=0,1,2,3$ can 
be used to
detect the curvature of the wave-front. Let us 
define $X_a^\mu=x_a^\mu-x_0^\mu$, the three baselines 
\cite{footd} connecting the
reference station ${\cal I}_0$ to the other three stations.
We write the components of $X_a^{\mu}$ as $(X_a,Y_a,Z_a,T_a)$.
Note that the baselines have spatial as well as 
temporal components. We now construct the following determinant
from these twelve numbers:
\begin{eqnarray}
S =\det \left\vert \begin{array}{ccc}
                      X_1 & X_2 & X_3\\
                      Y_1 & Y_2 & Y_3\\
		      (Z_1 - T_1) & (Z_2 - T_2) & (Z_3 -
		      T_3)\end{array}\right\vert
\label{determ}
\end{eqnarray}

If the wave-front were plane  with wave-vector 
$k^M_\mu=(\omega,0,0,-\omega)$,
the last row in the determinant $S$ would vanish
and so would $S$.  If the incident wave is plane, and has wave-vector
$k_\mu=(k_0,k_1,k_2,k_3)$, 
where $k_1,k_2$ are of order $\epsilon$, we find again that $S$ vanishes:
From $k_\mu k^\mu=k_0^2-k_1^2-k_2^2-k_3^2=0$, we conclude that to first 
order in $\epsilon$, $k_0=k_3$. It then follows from 
\begin{equation}
k.X_a=k_0(T_a-Z_a)-k_1 X_a-k_2 Y_a=0
\label{plane}
\end{equation}
that the last row of (\ref{determ}) is a linear combination of the first 
two and so $S$ vanishes. A non-vanishing $S$ implies that the wave-front
is not plane and therefore is a diagnostic for the curvature of the 
wave-front.

Theoretically the measured arrival time difference $T_a$
between stations $a$ and $0$  
can be written (to first order in $\epsilon$) as
\begin{equation}
T_a=Z_a-\tau_a
\label{expected}
\end{equation}
where $Z_a$ is the expected arrival time difference in the absence
of gravitational effects and $\tau_a=(\phi(x_a)-\phi(x_0))/\omega$. 
Plugging this
into the expression for $S$ we find that
\begin{equation}
\label{S}
S=(\tau_1 a_1+\tau_2 a_2 +\tau_3 a_3)
\end{equation}
where $a_1=X_2 Y_3-X_3 Y_2$ (and cyclic) are 
the areas of the parallelograms formed by the spatial baselines 
projected on the $x-y$ plane. 

Let us express $\tau_a$ using a Taylor expansion for $\phi(x)$
keeping terms up to second order in $x-x_0$, as in (\ref{taylor}).
It is convenient to choose co-ordinates so that the reference
event $x_0$ is in the $x-z$ plane, $x_0^y=0$. Setting 
$x_0^x=b$, the impact parameter, we find that
\begin{equation}
\tau_a=(1+\gamma)\rho[X_a/b+(Y_a^2-X_a^2)/b^2]+...
\label{Taylortau}
\end{equation}
As expected, the first term, being linear in $X_a$
drops out of the expression for $S$ and $S$ is given by
\begin{equation}
\frac{(1+\gamma)\rho}{b^2}[a_1(Y_1^2-X_1^2)+a_2(Y_2^2-X_2^2)+a_3(Y_3^2-X_3^2)]
\label{Sfinal}
\end{equation}

A non-vanishing $S$ signals curvature of the wave-front.
We can express the effect in time units. 
Given all the measured quantities $x_a^{\mu}$ except
$t_3$, if the wave-front were plane we would expect that 
the arrival time of the wave-front at station three would be
$t_3^e$ where $t_3^e$ satisfies the equation
\begin{equation}
(Z_1-T_1)a_1+(Z_2-T_2)a_2+(Z_3-t_3^e+t_0)a_3=0,
\label{false}
\end{equation}
 This expectation would be false, because $S$
does not vanish and we have
\begin{equation}
(Z_1-T_1)a_1+(Z_2-T_2)a_2+(Z_3-t_3+t_0)a_3=S
\label{true}
\end{equation} 
Subtracting (\ref{false}) from (\ref{true}) we find that
$t_3$ would differ from $t_3^e$ by an amount $\Delta t_3=t_3-t_3^e$
given by
\begin{equation}
\Delta t_3=S/a_3
\label{final}
\end{equation}
The difference $\Delta t_3$ between the measured
arrival time $t_3$ and the naive expectation $t_3^e$ would detect
the curvature of the wave-front.
The size of the effect is easily estimated 
from (\ref{S},\ref{final}). The main purpose of this letter is to draw 
attention
to the fact that this effect may be measureable with present 
VLBI technology.
Expressed in time units,
it is of order
\begin{equation}
10^{-5} |X|^2/b^2 s,
\label{estimate}
\end{equation} 
where $|X|$ is the typical size of the baseline (the intercontinental separation of the
telescopes) and $b$ the impact parameter. If one chooses 
the impact parameter to be three solar radii, ($b=3R_{\odot}$),
the effect is $100$ ps, which may be measureable.
At grazing incidence, ($b=R_{\odot}$), the effect is as large 
as a nanosecond. However,
lines of sight close to the Sun suffer from the problem of noise 
due to the Solar ionosphere.
As one chooses larger impact parameters, this noise is reduced, 
but so also is the
effect of interest (which decreases as the square of the impact parameter $b$).
The optimal choice of impact parameter may be best determined by trial and 
error. 

The ideal source for this experiment would be a strong, distant
point source.
A number of such sources have been already identified in the ICRF
(International Celestial Reference Frame) catalog \cite{ma}. 
These sources are densely distributed over the sky
with an average separation of a few degrees and 
can be detected using 
integration times of just 
a few minutes. Apart from the non-dispersive gravitational effects of
interest there are some dispersive non-gravitational effects 
such as due to ionospheric fluctuations. Such phases can be removed
by dual or multifrequency observations, which is a standard technique
in VLBI. There are also non-dispersive, non-gravitational effects
such as due to the troposphere. These can be removed by the
technique of phase referencing\cite{fomalont} between sources,
provided the sources 
are within a few degrees of each other.

As we mentioned earlier, the effect described here is contained in
the Shapiro effect, just as the bending of light is. Since the
Shapiro time delays are routinely accounted for in VLBI observations, 
the proposed effect has already been implicitly measured. The purpose
of this letter is to motivate observers to {\it explicitly} measure this
effect that general relativity predicts. Although the Shapiro effect has
been measured with high precision 
under varied conditions \cite{bertotti}, this does not constitute a {\it 
direct} 
measurement of the curvature of a wave front, since one does not 
sample four points on the same wavefront. We suggest here that a 
the curvature of the wave front is worth measuring directly.
Indeed, the mean curvature of the
wave-front is related via Raychaudhuri's equation\cite{wald} to the 
integral of the Ricci curvature along the line of sight. In Einstein's 
theory, we expect the wavefront to be a minimal surface (zero mean
curvature). This prediction could be explicitly checked.
To conclude, we have proposed an idealised experiment to directly measure 
the curvature of the wave-front from a distant radio source. We hope to
motivate VLBI astronomers to design a more realistic version of the 
experiment to measure the relativistic curvature of wave fronts.

{\it Acknowledgements:} It is a pleasure to thank K.G. Arun,
D.Bhattacharya, Abhishek Dhar, BR Iyer, N.Kumar, S. Nair, B. Nath, 
R. Nityananda, C.S. Shukre, Sukanya Sinha, Supurna Sinha  and C.R. 
Subrahmanya for their comments. I also acknowledge anonymous 
referees' comments which have much improved the paper.


\vbox{
\vspace{1.0cm}
\epsfxsize=6.0cm
\epsfysize=6.0cm
\begin{figure}
\caption{Figure 1 shows disposition of the radio telescopes,
all on the same hemisphere as the source S1, which is in common
view of all the telescopes. The source S2 about three degrees away from
S1 is used for phase referencing. The figure is not to scale.}
\label{sun}
\end{figure}}

\begin{references}
\vspace{.5cm}
\bibitem{fomalont}
Fomalont E and Kopeikin S {\it Proc. of the 6th international VLBI
symposium}, Eds. Ros E et al (2002);
Kopeikin S and Fomalont E, eprint gr-qc/0206022.   
\bibitem{sovers}
Sovers OJ {\it et. al.} , {\it Rev. Mod. Phys.} {\bf 70}, 1393 
(1998).
\bibitem{shapiro}
Shapiro II {\it Phys. Rev. Lett.} {\bf 13}, 789 (1964).
\bibitem{will}
Will CM,{\it Theory and Experiment in Gravitational Physics}
(Cambridge:CUP) 1993.
\bibitem{will1}
Will CM {\it Liv. Rev. Relativ.}, 4, (2001-4);
Eubanks TM et al, {\it Advances in solar system tests of gravity}, 
(August, 1999), [Online Preprint]: cited on 15
January 2001, ftp://casa.usno.navy.mil/navnet/postscript/prd\_15.ps;
Lebach DE et al, {\it Phys. Rev. Lett.} {\bf 75}, 1439 (1995).
\bibitem{scalar}
We treat the electromagnetic waves as scalar waves by restricting
attention to a single polarisation.
\bibitem{landau}
Landau LD and Lifshitz EM,  {\it Classical Theory of Fields},
Fourth Edition, Pergammon Press (1975).
\bibitem{curve}
Theoretically the curved wave-front has principal curvatures which
are equal and opposite. Thus, the wave-front is a minimal surface
with negative Gaussian curvature. 
\bibitem{astro}
We are referring here to the curvature of an incident plane wave-front
due to the Sun's gravitational field. We are not talking about the 
curvature of the wave-fronts \cite{sovers} of radiation {\it emitted by 
bodies at a finite distance}. 
\bibitem{footd}
We suppose the locations of the stations so chosen that of the
four three dimensional vectors ${\bf X}_1,{\bf X}_2,{\bf X}_3,{\bf k}$
every triple has a triple product well away from zero. This will prevent
one station ``shadowing'' another and linearly dependent baselines. 
\bibitem{ma}
Ma et al, {\it Astronomical Journal}, {\bf 116}, 516 (1998).
\bibitem{bertotti} Bertotti et al, {\it Nature} {\bf 425}, 374 (2003).
\bibitem{wald}
Wald RM, {\it General Relativity}, University of Chicago Press, (1984).
\end{references}
\end{document}